\documentclass[letterpaper, conference]{IEEEtran}
\IEEEoverridecommandlockouts
\usepackage{tablefootnote} % for table footnotes
\usepackage{threeparttable}
\usepackage{cite}
\usepackage{svg}
\usepackage{amsmath}
\usepackage{cuted}
\usepackage{amssymb,amsfonts, epsfig,latexsym}
\usepackage{algorithm}
\usepackage{algpseudocode}

\usepackage{subfigure}
\usepackage{flushend}
\usepackage{soul}
\usepackage{hhline}
\usepackage{colortbl}
\usepackage{xcolor}
\usepackage{pdfpages}
\usepackage{graphicx}
\usepackage{textcomp}

\usepackage{multirow}
\usepackage{bigstrut}
\usepackage{array}
\usepackage{comment}
\usepackage{booktabs}
\usepackage{multicol}
\usepackage[utf8]{inputenc}
\renewcommand{\arraystretch}{1.5}

\def\BibTeX{{\rm B\kern-.05em{\sc i\kern-.025em b}\kern-.08em
    T\kern-.1667em\lower.7ex\hbox{E}\kern-.125emX}}

\usepackage[none]{hyphenat}
\usepackage[hyphens]{url}
\usepackage[textsize=tiny,textwidth=0.72in]{todonotes}
\usepackage{color} % color added for editing
\usepackage{soul}

\begin{document}

%\title{Custom HLS and HDL Design of LSTM Accelerator for High-Rate Dynamic Applications\\}
%\title{Accelerating High-Rate Dynamic Systems with LSTM Networks\\}
% JDB
%\title{Low Latency Attention Block of Transformer Encoder on FPGA\\}
%\title{FLAME: Flexible Accelerator for the Attention Mechanism of Transformer on FPGA\\}
%\title{FLARE: Flexible Accelerator for the Attention Mechanism in Transformers on FPGA\\}
%\title{FLOAT: Flexible Accelerator for the Attention Mechanism in Transformers on FPGA\\}
%\title{FORAM: Flexible Accelerator for the Attention Mechanism in Transformers on FPGA\\}
%\title{FAMe: Flexible Accelerator for the Attention Mechanism in Transformers on FPGA\\}
%\title{RATE: Runtime Programmable Accelerator for Transformer Encoder on FPGAs\\}
\title{ProTEA: \textbf{\underline{Pro}}grammable \textbf{\underline{T}}ransformer \textbf{\underline{E}}ncoder \textbf{\underline{A}}cceleration on FPGA\\}
%\author{\IEEEauthorblockN{Ehsan Kabir$^{\mathrm{*}}$, David Andrews$^{\mathrm{*}}$, Miaoqing Huang$^{\mathrm{*}}$}

%\IEEEauthorblockA{\textit{$^{\mathrm{*}}$Department of Computer Science and Computer Engineering, University of Arkansas, USA} \\  
%\{ekabir, dandrews, mqhuang\}@uark.edu}{\centering}}

\author{\IEEEauthorblockN{Ehsan Kabir$^{\mathrm{*}}$, Jason D. Bakos$^{\mathrm{\dag}}$, David Andrews$^{\mathrm{*}}$, Miaoqing Huang$^{\mathrm{*}}$}
\IEEEauthorblockA{\textit{$^{\mathrm{*}}$Department of Electrical Engineering and Computer Science, University of Arkansas, Fayetteville, USA} \\
\textit{$^{\mathrm{\dag}}$Department of Computer Science and Engineering, University of South Carolina, USA} \\ 
%\ { \\ }{\centering}
\{ekabir, dandrews, mqhuang\}@uark.edu, jbakos@cse.sc.edu}}{\centering}

%\author{\IEEEauthorblockN{}

%\IEEEauthorblockA{\textit{} \\
%\textit{}\\
%\textit{}\\
%}}{\centering}

%\and
%\and
%\and
%\and
%\and
%\IEEEoverridecommandlockouts
%\IEEEpubid{\makebox[\columnwidth]{978-1-5386-5541-2/18/\$31.00~\copyright2018 IEEE \hfill}
%\hspace{\columnsep}\makebox[\columnwidth]{ }}

\onecolumn
{\large\vspace*{\fill}

© 2024 IEEE.  Personal use of this material is permitted.  
Permission from IEEE must be obtained for all other uses, in any current or future media, including reprinting/republishing this  material for advertising or promotional purposes, creating new collective works, for resale or redistribution to servers or lists, or reuse of any copyrighted component of this work in other works. \\

This work has been accepted at the SC24 (The International Conference for High-Performance Computing, Networking, Storage, and Analysis) Workshop Proceedings and will appear in the proceedings and on the IEEE website soon.

% Create vertical space to center the paragraph
\vspace*{\fill}
}
\twocolumn

\maketitle

%\IEEEpubidadjcol

\begin{abstract}
Transformer neural networks (TNN) have been widely utilized on a diverse range of applications, including natural language processing (NLP), machine translation, and computer vision (CV). Their widespread adoption has been primarily driven by the exceptional performance of their multi-head self-attention block used to extract key features from sequential data. The multi-head self-attention block is followed by feedforward neural networks, which play a crucial role in introducing non-linearity to assist the model in learning complex patterns. Despite the popularity of TNNs, there has been limited numbers of hardware accelerators targeting these two critical blocks. Most prior works have concentrated on sparse architectures that are not flexible for popular TNN variants. This paper introduces \textit{ProTEA}, a runtime programmable accelerator tailored for the dense computations of most of state-of-the-art transformer encoders. \textit{ProTEA} is designed to reduce latency by maximizing parallelism. We introduce an efficient tiling of large matrices that can distribute memory and computing resources across different hardware components within the FPGA.  We provide run time evaluations of \textit{ProTEA} on a Xilinx Alveo U55C high-performance data center accelerator card.  Experimental results demonstrate that \textit{ProTEA} can host a wide range of popular transformer networks and achieve near optimal performance with a tile size of 64 in the multi-head self-attention block and 6 in the feedforward networks block when configured with 8 parallel attention heads, 12 layers, and an embedding dimension of 768 on the U55C. Comparative results are provided showing \textit{ProTEA} is 2.5$\times$ faster than an NVIDIA Titan XP GPU. Results also show that it achieves 1.3 -- 2.8$\times$ speed up compared with current state-of-the-art custom designed FPGA accelerators.
% which CPU and GPU ?   6.5$\times$ faster than JETSON TX2 GPU and
% and the proposed architecture achieves ...×, ....× improvement in latency, and ...×, ...× energy savings compared to CPU and GPU. and 1.29 $\times$

\end{abstract}


\section{Introduction}
In recent years, transformer neural networks have become widely utilized on a diverse range of applications including natural language processing (NLP) \cite{language, attention}, neural machine translation \cite{NMT}, and image processing \cite{wang_via_2022}. They are becoming favored over traditional recurrent neural network (RNN) and long short-term memory (LSTM) models for NLP tasks, and convolutional neural networks (CNN) for CV tasks. Their popularity is being driven by their ability to enable high computational parallelism for both the training and inference steps. Their natural exposure of higher levels of parallelism makes them well-suited for acceleration on hardware such as GPUs and FPGAs. There exist many transformer-based models such as full transformers containing both encoder and decoder \cite{attention}, BERT \cite{bert}, RoBERTa \cite{robert}, Swin Transformers \cite{liu_efficient_2023}, structBERT \cite{structbert} etc. These models incorporate two notable features: a multi-headed attention (MHA) mechanism and feedforward neural networks (FFN) that distinguishes them from traditional CNNs, RNNs, and LSTMs.
These MHA and FFN mechanisms are computationally expensive due to intensive matrix-matrix multiplications and complex data flows \cite{sanger}. They account for a significant portion of runtime in many existing TNNs \cite{ham_elsa_2021}. Unfortunately, executing TNNs is inefficient on general-purpose platforms such as GPUs and CPUs because of their high power consumption, low computational efficiency, underutilized memory bandwidth, and significant compilation overheads\cite{flightLLM}. 
In addition to GPUs, FPGAs have become popular commercial off the shelf components used to accelerate DNNs. FPGAs offer the ability to exploit high level of parallelism to provide low run time inference latencies with efficient power consumption \cite{dl, gan_hardware-aware_2022}. Many studies have investigated how to increase the parallelization of CNNs, LSTMs, Graph Convolutional Networks \cite{cnn_low_batch, lstm_high_rate, gcn_large, que_optimizing_2022} on FPGAs to enhance performance.  Recently, TNNs have been successfully deployed on FPGAs and application-specific integrated circuit (ASIC) hardware accelerators\cite{lu_hardware_2020, A3AA, peng_length_2022}. Most implementations compress the model by using different weight pruning strategies, and reduce latency by incorporating sparse matrices. Thus, they use a specialized sparse architecture specific to each application. However, different applications require different sparsity patterns, necessitating the redesign of the hardware architecture for optimal results. This comes at the cost of time-consuming synthesis, and requires skills in digital design and computer architecture as well as detailed knowledge of each target logic family.  %This is similar to the limitations of ASICs which are also designed for a fixed configuration, preventing the expensive chip from being used for different models or the same model with varying configurations\cite{FaFlA}.  
%Custom FPGA accelerators often lack the ability to be reconfigured for different models during runtime. 
Therefore, there is a need for a versatile accelerator capable of efficiently managing dense matrix computations across a range of TNN applications.

The study in \cite{lu_hardware_2020} uses logic resources to implement a systolic array for parallelism, which can lead to underutilization of digital signal processing (DSP) units that are capable of high-speed computation at higher frequencies. DSP utilization also depends on the implementation method. For instance, many accelerators \cite{peng_length_2022, peng_accelerating_2021, jiang_ultra_nodate, wojcicki_accelerating_2022} employ high-level synthesis (HLS) tools, while others use hardware description language (HDL) \cite{chen_high-frequency_2023, yang_efa-trans_2022, bai_ltrans-opu_nodate} for design. Although HLS requires less implementation time compared to HDL, writing efficient HLS code that effectively manages specific FPGA resources, such as DSPs, for optimal performance remains challenging \cite{lstm_high_rate}. %Consequently, the application can only operate at a low clock frequency, increasing the chance of timing violation issues due to the high resource utilization. The attention mechanism also involves large matrices. 

The analysis in \cite{mha_compute, qi_accelerating_2021, li_ftrans_2020, luo_calabash_2023} demonstrated that MHA and FFN occupy major portions of the memory and they have the highest computational demands. %For example, the largest BERT (bidirectional encoder representations from transformers) model can have a matrix size of 512×768, which requires more than 20MB to store all the necessary data of a single layer, according to Calabash\cite{luo_calabash_2023}.
Since on-chip memory of FPGAs typically does not exceed 36MB and off-chip memory bandwidth is sometimes limited, matrices must be partitioned into tiles. However, designing an optimal partitioning scheme for MHA and FFN that aligns effectively with the architecture presents a significant challenge.

In this paper, HLS tool was used to design \textbf{\textit{ProTEA}}, a programmable accelerator for transformer encoders. The code of the design written in HLS was optimized to increase the parallel computations by the DSPs. % in parallel. %The computations are performed by DSP48 slices, while the data used by the DSPs is stored in BRAMs, LUTRAMs, or registers within the processing elements (PE). %Literature [12] presents an effective single-instruction multiple-data (SIMD) 3-dimension revolution engine designed for FPGA through mapping two multiply and accumulate (MAC) operations within a single computing unit (e.g., DSP slice). Therefore, we introduce double MAC (DMAC) ideas to SA and optimize them simultaneously. To the best of our knowledge, this is the first architecture that efficiently applies DSP48 slices for SA to accelerate the attention and support dynamic reconfiguration of SA to adapt to input data with different bit widths. Despite significant advancements in related domains, these potent transformer networks' high computational complexity and large memory needs make them difficult to function in embedded systems or mobile devices. This issue is drawing the attention of researchers.
%In this paper, we mapped independent processing elements (PE) the HLS is used to design the accelerator where 
%Each PE has dedicated DSPs, BRAMs and registers as on-chip storage. 
\textbf{\textit{ProTEA}} incorporates efficient tiling for both the attention mechanism and linear transformations. It ensures enhanced parallel computations and communication so that the transformer encoding can be accelerated as much as possible. %introduce double MAC (DMAC) ideas to SA and optimize them simultaneously. To the best of our knowledge, this is the first architecture that efficiently applies DSP48 slices for SA to accelerate the attention and support dynamic reconfiguration of SA to adapt to input data with different bit widths. In spite of making great progress in relative fields, the high computation complexity and huge memory requirements of these powerful Transformer networks are making them hard to be operated in mobile devices or embedded systems. More and more researchers are paying attention to this problem.
%\hfill \\
%\hfill \\

\textit{The contributions of this paper are:}
%\vspace{-3.5mm}
\begin{itemize}
    %\item [$\bullet$] An accelerator for the encoder of transformer that achieves low latency by exploiting the available parallelism of FPGA.  
    
    \item [$\bullet$] A novel accelerator architecture for transformer encoders that maximizes DSP utilization to enhance parallel processing and achieve low latency.
    
    \item [$\bullet$] An efficient tiling strategy for weight matrices in both the multi-head attention layer and the feedforward neural network layer, enabling the accommodation of large models within on-chip memory.
    
    \item [$\bullet$]  A parameterized HLS code that allows for design-time adjustments of parameters in the HLS tool.% such as the number of attention heads, tile size, embedding dimension, bit width, and sequence length.

    \item [$\bullet$] A runtime programmable feature enabling dynamic adjustment of parameters in software, %like the number of attention heads and encoder layers, embedding dimension, and sequence length, 
    facilitating the evaluation of different models without the need for hardware re-synthesis.
    
\end{itemize}

\section{Background}\label{back}

\begin{figure}[h]
\centering
    %\begin{subfigure}{.5\textwidth}
    \centering
    \includegraphics[height=10cm, width=0.9\linewidth]{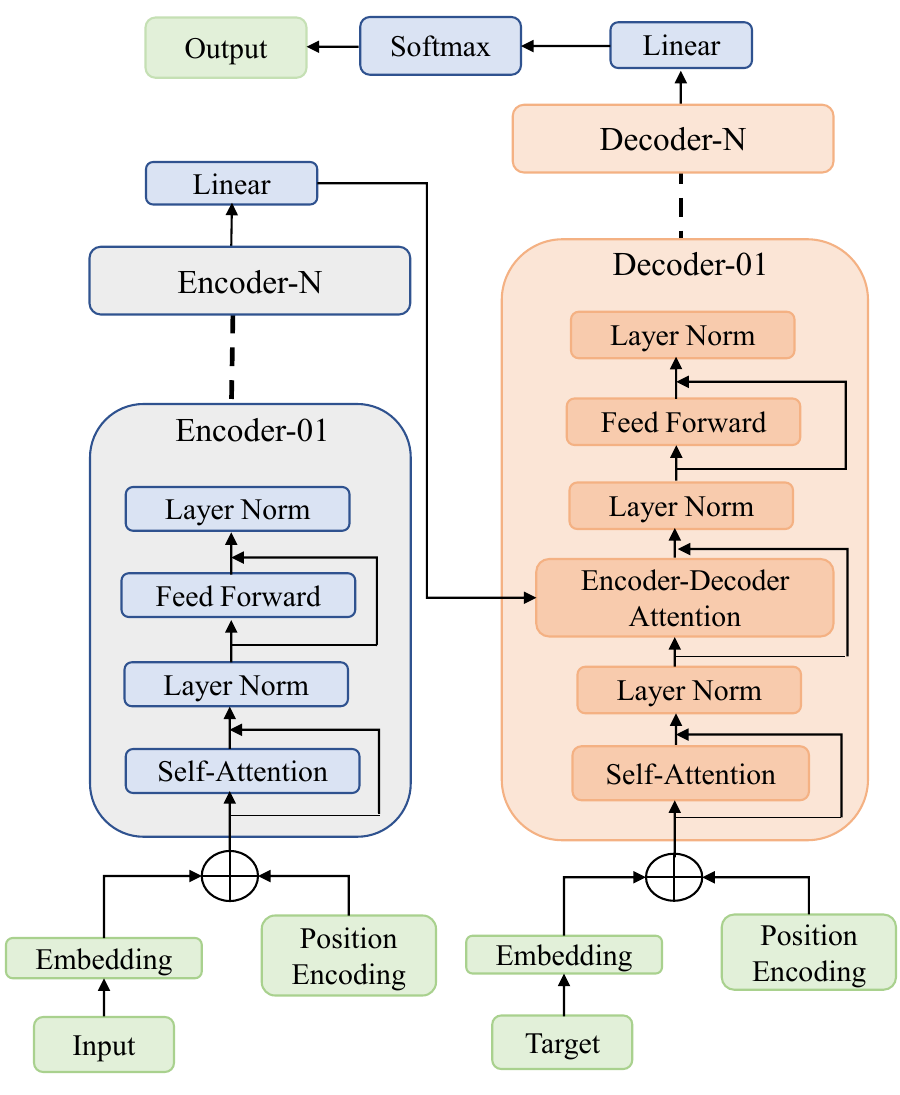}
    \caption{\label{TNN}Transformer Architecture.}
\end{figure}

Transformers consist of several fundamental components, as depicted in Fig.~\ref{TNN}. An input sequence of tokens is first converted into embeddings. The positional encoder adds positional information to these embeddings, enabling the model to account for the order of tokens in a sequence. This encoder generates vectors that provide context based on each word's position in a sentence. These vectors are then linearly transformed into three tensors: Q (queries), K (keys), and V (values) by multiplying the embedding matrix with three distinct weight matrices. The encoder block processes these tensors, transforming them into a higher-level representation that captures essential information. This transformation is crucial for accurately capturing features and contextual relationships within the input sequence. The encoder architecture is composed of two primary sub-layers: (1) the self-attention mechanism, and (2) the position-wise feed-forward network.

The self-attention mechanism allows the model to simultaneously evaluate different parts of an input sequence, capturing long-range relationships by calculating attention scores and using multi-head projections for various input representations. This capability enables the model to effectively learn complex patterns, dependencies, and relationships. The position-wise feed-forward network (FFN), similar to a multilayer perceptron (MLP), applies linear transformations independently to each position in the input sequence. This network performs two linear transformations, primarily involving matrix-vector multiplication. The first transformation includes activation functions such as the Rectified Linear Unit (ReLU) or Gaussian Error Linear Unit (GeLU), while the second transformation does not.

Additionally, each sub-layer incorporates a residual connection combined with layer normalization (LN), addressing the vanishing gradient problem during training. Residual connections and LN layers are added after each MHA and FFN layer, involving the addition of matrix elements and nonlinear functions.

The decoder block, depicted in Fig.~\ref{TNN}, is tasked with generating the output sequence using the encoded representations provided by the encoder. Similar to the encoder, the decoder comprises a stack of N identical layers. Each layer in the decoder includes three sub-layers: (1) the Masked Attention Mechanism, which is similar to the encoder's self-attention but incorporates a masking feature to prevent the output from depending on future outputs; (2) an attention layer that focuses on the encoder's output, allowing the decoder to highlight relevant parts of the input sequence for each output element; and (3) a position-wise feed-forward network.

\begin{figure}[h]
\centering
    %\begin{subfigure}{.5\textwidth}
    \centering
    \includegraphics[height=6cm, width=1.0\linewidth]{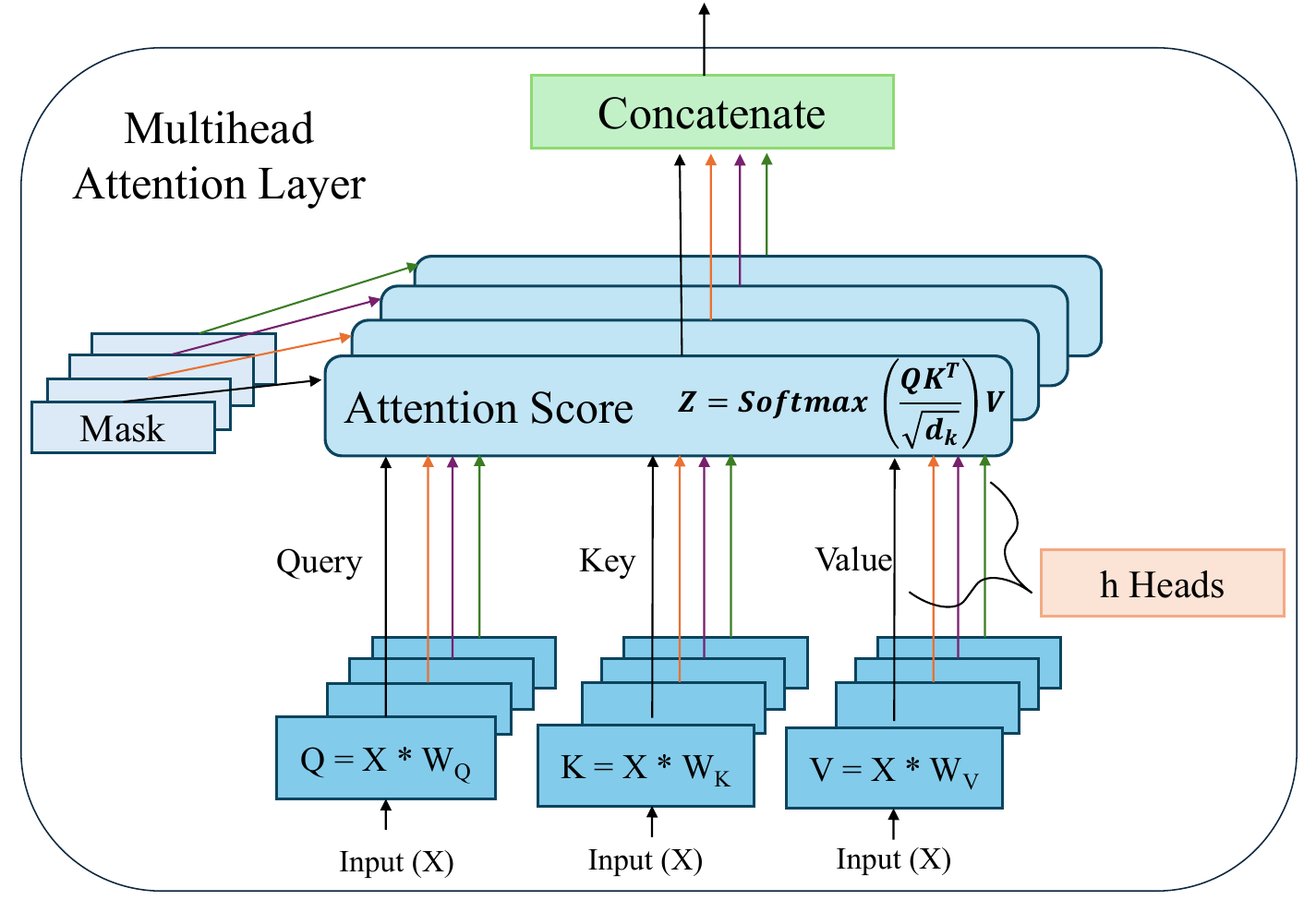}
    \caption{\label{mha}Multihead Attention Layer.}
\end{figure}

As shown in Fig~\ref{mha}, the scaled dot-product attention in each head is a vital component of the multi-head attention layer. The attention weights are calculated by taking the dot product of the Q and K matrices and then scaling the result by the square root of the second dimension of the K matrix. This scaling is crucial to prevent the dot products from becoming too large, which helps stabilize gradients during training. The scaled dot products are then passed through the softmax function to compute the attention weights. These weights are used to perform a weighted sum of the value vectors. The final output is the projection of the concatenated sequences from all heads.

The output of MHA can be represented as Equation 1 \& 2. The input sequence X is linearly mapped into $Q_i, K_i, V_i$ matrices using weights and biases. The parameter $d_k = d_{model}/h$ is the 2\textsuperscript{nd} dimension of $Q_i$ and $K_i$. $d_{model}$ is a hyperparameter called embedding dimension and h is number of heads. `i' is the index for attention heads.

\begin{equation}      %\resizebox{.8\hsize}{!}
\label{eq:attention}
    Attention (Q_i, K_i, V_i) = softmax \left( Mask \left(\frac{Q_iK^T_i}{\sqrt{d_k}} \right)\right)V_i
\end{equation}

\begin{align}
\begin{split}
    Q_i = X \times W_q + B_q, \\K_i = X \times W_k + B_k, \\V_i = X \times W_v + B_v
\end{split}
\end{align}
\hfill
%\begin{equation}
    %\resizebox{.8\hsize}{!} 
%    Q_i = X\times W_q + B_q, K_i = X\times W_k+ B_k, V_i = X\times W_v + B_v
%\end{equation}

%\input{dynamics}
%\input{model}
\section{Related work}\label{work}
Various FPGA and ASIC accelerators have been designed for TNNs. The ASIC design in \cite{A3AA} leveraged parallelism and specialized datapaths to achieve significant gains in performance and energy efficiency. Another ASIC, ELSA \cite{ham_elsa_2021}, employed specialized approximation algorithms to reduce computational demands. The SpAtten \cite{spattn} ASIC utilized sparsity and quantization to decrease computations and memory access. Additionally, the hardware-software co-design framework Sanger \cite{sanger} facilitated dynamic sparsity through a reconfigurable ASIC architecture. Despite these advancements, these solutions primarily focus on accelerating sparse attention mechanisms and do not address the deployment of full transformer models. The FPGA accelerator proposed by Lu et al. \cite{lu_hardware_2020} is the first hardware architecture to accelerate both the MHA and FFN layers of the transformer. However, their implementation was done using HDL for a single attention head. %Ye et al.\cite{ye_accelerating_2023} proposed an FPGA accelerator for MHA with reconfigurable architecture, efficient systolic arrays, and hardware-friendly radix-2 softmax, but they did not consider maximization of BRAM and DSP usage to maximize parallelism. Fujimaki et al. \cite{fujimaki_self-attention_2022} also proposed a systolic array-based accelerator for attention mechanism within CNN where MHA consumed 63\% of the computation time, but they used MHA within CNN, not TNN.
A shared computing architecture is implemented in \cite{qiao_fpga-based_2022}, where a parallel computing array is shared between MHA and FFNs for a CNN application. A novel structural pruning method was proposed by \cite{zhang_algorithm-hardware_2021} and the associated accelerator on FPGA was designed to reduce memory footprint. Peng et al. \cite{peng_accelerating_2021} explored column-balanced block-wise pruning for transformers and designed an FPGA accelerator for optimized block-wise matrix multiplication. An algorithm hardware framework \cite{qi_accelerating_2021} utilizes latency and accuracy constraints to determine the optimal sparsity ratio and select an appropriate FPGA platform. The energy-efficient acceleration framework FTRANS \cite{li_ftrans_2020} features an improved block-circulant matrix method for algorithm-level sparsity, along with a custom-designed accelerator tailored for this approach. Wojcicki et al.\cite{wojcicki_accelerating_2022} deployed a small TNN model on FPGA using HLS for experiments at the Large Hadron Collider. All of the existing hardware architectures are designed for a specific TNN and a specific sparsity pattern. They lack the flexibility to reconfigure the computing structure for different applications during runtime. EFA-Trans \cite{yang_efa-trans_2022} is compatible with dense and sparse computing patterns, but it would need resynthesis of the hardware to switch between two options. Furthermore, none of them explored which tile size and what utilization DSPs could achieve optimum parallelism. 
\section{Accelerator Architecture}\label{arch}
\begin{figure*}[h!]
\centering
\includegraphics[height=6cm, width=0.7\linewidth]{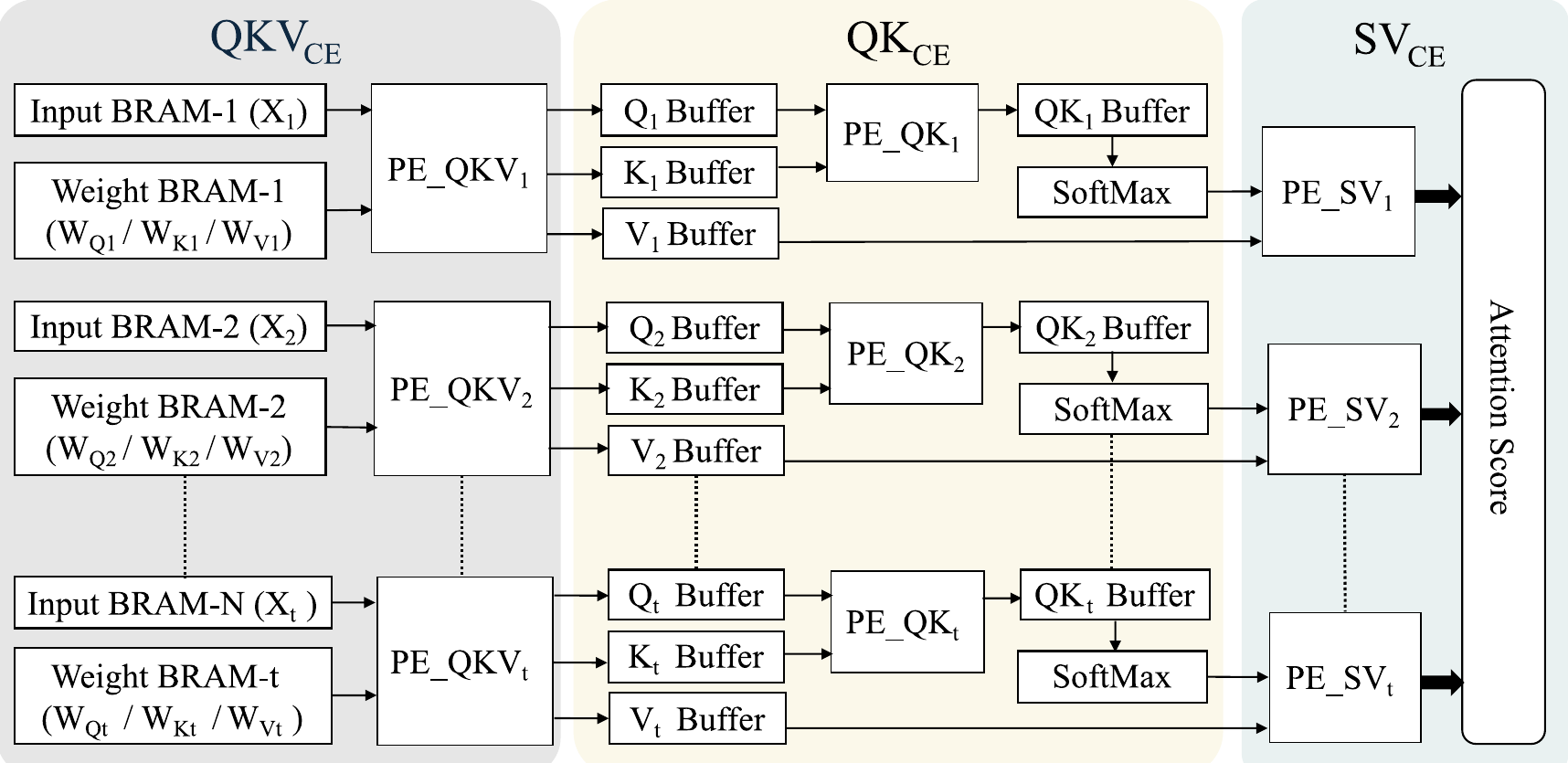} % 6cm 0.7cm
\caption{\label{mha_System}Computations of the Attention Module}
\end{figure*}

The core of the accelerator is designed in C language on Vitis HLS 2022.2.1 tool. C simulation verifies the correctness of the algorithm, while C/RTL co-simulation ensures the functionality of the synthesized hardware. This section describes the high-level synthesis design technique that generates an optimized architecture utilizing most of the DSPs in the computation engines, ensuring high parallelism. %\subsection{Overall Structure}
The overall structure of the accelerator contains two main processing modules - the multihead attention  (MHA) module and the feedforward network (FFN) module, which are shown in Fig.~\ref{mha_System} and Fig.~\ref{FF_System} respectively. The overall system was designed in Vivado 2022.1.2 design suite. It contains a custom IP block for the accelerator, which is exported from HLS. The inputs and weights are fetched from off-chip high-bandwidth memory (HBM) using AXI4 master interfaces \cite{axiMaster} when the load instruction from the accelerator controller is received according to demand. The accelerator receives control signals from the processor through an AXI-lite slave interface \cite{axiSlave}. Each hyperparameters of TNN can be programmed during runtime up to a maximum value by MicroBlaze ($\mu$B) softcore processor \cite{microblaze}.

\subsection{Attention Module}\label{am}
The attention module (Fig.~\ref{mha_System}) comprises three computation engines (CE), labeled as ${QKV}_{CE}$, ${QK}_{CE}$, and ${SV}_{CE}$ based on their outputs. The number of these engines is determined by the number of attention heads (h). Each engine features an array of processing elements (PE), where each PE includes a DSP48 for performing multiplication and accumulation (MAC) operations. The quantity of PEs denoted as `t' is influenced by the unrolling factor of the inner loop and the initiation interval of the pipelined outer loop. Since the data access patterns and computational needs vary across different engines, each has separate function definition in HLS. This ensures that the synthesized RTL modules of the engines contain distinct PE arrays, enabling individual optimization. Input data and weights are stored in multiple BRAMs/LUTRAMs to support parallel access.

Each PE operates independently, equipped with its own memories, controller, and computing units. In HLS, the weights ($W_q$, $W_k$, $W_v$) for generating the query (Q), key (K), and value (V) matrices are defined as separate two-dimensional arrays of size ($\frac{d_{model}}{h} \times TS_{MHA}$). Here, $TS_{MHA}$ represents the tile size in the attention module. It is the dimension of the sub-matrices into which the larger weight matrices are partitioned. The number of heads, tile size, and array partitioning directives in HLS determine how these arrays are divided to create multiple two-port BRAMs. To address the limited ports of BRAMs, array partitioning and data loading are optimized to ensure that data needed simultaneously by a DSP is stored in separate BRAMs. The Q, K, and V matrices, sized ($SL \times \frac{d_{model}}{h}$), are stored in intermediate buffers. Here, SL stands for sequence length. %, which are also implemented using BRAMs

%\vspace{0.1cm}
\subsubsection{\textbf{QKV\textsubscript{CE} engine}}
${QKV}_{CE}$ engine generates the query, key, and value matrices. This engine contains the $W_q$, $W_k$, $W_v$ buffers, and input ($X_i$) buffers from which data is accessed in parallel by parallel DSP units. The arrays used in this engine are divided into subarrays using our tiling technique to fit into on-chip memories. %Experimental results demonstrated that $TS_{MHA}$ of 64 is optimal for HLS, allowing for efficient array partitioning within a reasonable compilation time (approximately 36 hours) for a state-of-the-art (SOTA) transformer encoder. %The influence of tile size on performance is detailed in Section~\ref{results}.
The number of loop iterations in the ${QKV}_{CE}$ engine is determined by $TS_{MHA}$, resulting in a total of ($\frac{d_{model}}{TS_{MHA}}$) tiles or iterations. During each iteration, distinct data is loaded into the $W_q$, $W_k$, $W_v$, and $X_i$ buffers. Computations then commence in the PEs, while biases for the Q, K, and V matrices are simultaneously loaded into registers from off-chip memory. These biases are subsequently added to the Q, K, and V matrices. Algorithm~\ref{QKV} illustrates the computations of this engine, where the second loop (line 6) is pipelined, resulting in the full unrolling of the innermost loop (line 8) and generating ($\frac{d_{model}}{TS_{MHA}}$) PEs.

%\vspace{0.1cm}
\subsubsection{\textbf{QK\textsubscript{CE}\ engine}}
The ${QK}_{CE}$ engine performs matrix-matrix multiplication between the Q and K matrices. Since these matrices are relatively small, they are not tiled. Algorithm~\ref{QK} outlines the operations performed.
%\vspace{-0.8cm}
\begin{algorithm}[H]
\centering
\caption{Q, K, V Calculation}\label{QKV}
\begin{algorithmic}[1]
%\State Inputs
\For{$i \gets 1$ to $Sequence\ Length$}
\State \textbf{\#pragma HLS pipeline off}
\State $S_q \gets 0$
\State $S_k \gets 0$
\State $S_v \gets 0$
\For{$k \gets 1$ to $\frac{Embedding\ Dimension}{Number\ of\ Heads}$}
\State \textbf{\#pragma HLS pipeline II = 1}
\For{$j \gets 1$ to $Tiles\ in\ MHA$}
    \State $S_q \gets S_q + x[i][j]\times w_q[k][j];$
    \State $S_q \gets S_q + x[i][j]\times w_k[k][j];$
    \State $S_q \gets S_q + x[i][j]\times w_v[k][j];$
\EndFor
    \State $Q[i][k] \gets Q[i][k] + S_q;$
    \State $K[i][k] \gets K[i][k] + S_k;$
    \State $V[i][k] \gets V[i][k] + S_v;$
\EndFor
\EndFor
\end{algorithmic}
\end{algorithm}
The innermost loop (line 6) is fully unrolled, resulting in ($\frac{d_{model}}{h}$) PEs for this engine. %This module includes BRAMs for Q and K matrices, from which data is accessed by the DSP units. The division operation described in Equation \ref{eq:attention} is executed within this module using lookup tables (LUTs), with parallel operations limited to avoid excessive LUT utilization. 
The engine generates a matrix (S) of attention weights, which is stored in either BRAM or registers. These values are then passed to the softmax function. The softmax function, implemented in HLS, utilizes LUTs and flip-flops (FFs) to compute the result.

%\vspace{-0.6cm}
\begin{algorithm}[H]
\centering
\caption{$Q\times K^T$ Calculation}\label{QK}
\begin{algorithmic}[1]
%\State Inputs
\For{$i \gets 1$ to $Sequence\ Length$}
\State \textbf{\#pragma HLS pipeline off}
\For{$j \gets 1$ to $Sequence\ Length$}
\State \textbf{\#pragma HLS pipeline II = 1}
\State $S \gets 0$
\For{$k \gets 1$ to $\frac{Embedding\ Dimension}{Number\ of\ Heads}$}
    \State $S \gets S + Q[i][k]\times K[j][k];$
\EndFor
    \State $s[i][j] \gets S/Embedding\_Dimension;$

\EndFor
\EndFor
\end{algorithmic}
\end{algorithm}

%\vspace{0.1cm}
\subsubsection{\textbf{SV\textsubscript{CE} engine}}
The output matrix (S) from the softmax operation is passed to the ${SV}_{CE}$ engine (Algorithm~\ref{SV}), where it undergoes matrix-matrix multiplication with the value (V) matrix. In Algorithm~\ref{SV}, the innermost loop (line 6) is fully unrolled, resulting in (SL) PEs. The output from this engine is termed the attention score.%Scores are generated for each head, which are subsequently concatenated. %($\frac{d_{model}}{h}$)
%\vspace{0.4mm}
%within a separate module (not shown in figure). The number of BRAMs and DSPs depends on the tile size and number of heads defined during design time. 
%\vspace{-0.3cm}
\begin{algorithm}[h]
\centering
\caption{$S\times V$ Calculation}\label{SV}
\begin{algorithmic}[1]
%\State Inputs
\For{$i \gets 1$ to $Sequence\ Length$}
\State \textbf{\#pragma HLS pipeline off}
\For{$j \gets 1$ to $\frac{Embedding\ Dimension}{Number\ of\ Heads}$}
\State \textbf{\#pragma HLS pipeline II = 1}
\State $vv \gets 0$
\For{$k \gets 1$ to $Sequence\ Length$}
    \State $vv \gets vv + S[i][k]\times V[k][j];$
\EndFor
    \State $SV[i][j] \gets vv;$
\EndFor
\EndFor
\end{algorithmic}
\end{algorithm}

%\vspace{-0.4cm}
\subsection{Feedforward Network Module}
\begin{figure*}[h]
\centering
\includegraphics[height=5cm, width=1.0\linewidth]{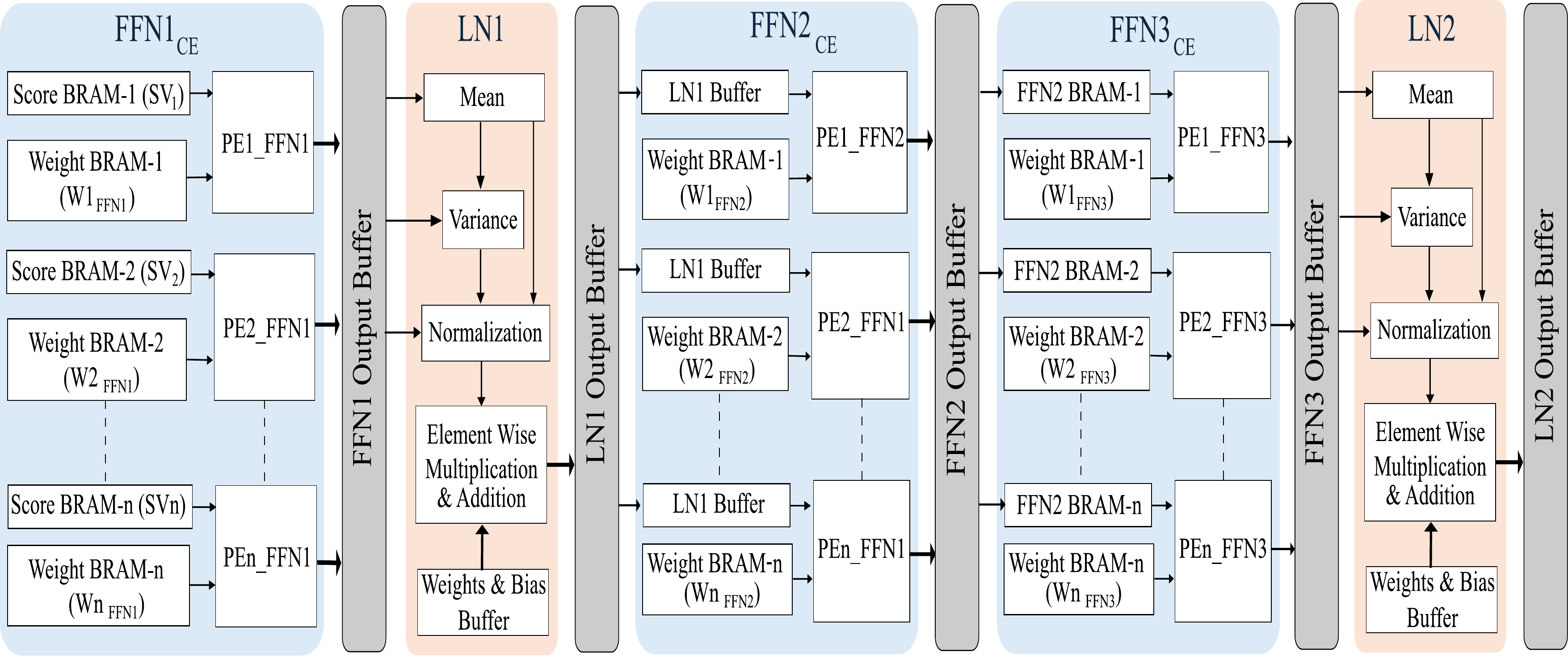} % 6cm 0.7cm
\caption{\label{FF_System}Computations of Feedforward Network.}
\end{figure*}

There are three CEs, denoted as $FFN1_{CE}$, $FFN2_{CE}$, and $FFN3_{CE}$ in FFN to perform the operations of feedforward networks of different dimensions (Fig.~\ref{FF_System}). The definitions of the functions representing the CEs have different dimensions of arrays for the inputs and outputs in HLS. These arrays are converted into BRAMs/LUTRAMs after synthesis. The number of computations inside each engine is different, which is why each has a separate function in HLS. They contain a different number of processing elements (PE) after synthesis because of different unrolling factors of the innermost loop. The weights are stored in a two-dimensional array ($W_o$) of size  ($\frac{d_{model}}{TS_{FFN}} \times \frac{(4\times d_{model})}{TS_{FFN}}$) in HLS, where $TS_{FFN}$ is tile size in FFN. $FFN1_{CE}$ and $FFN3_{CE}$ are followed by layer normalization (LN) modules. Algorithm~\ref{FFN1} describes the general coding approach for an FFN engine.
%\vspace{0.2cm}
\subsubsection{\textbf{{FFN1}\textsubscript{CE} engine}}
$FFN1_{CE}$ engine performs the first linear transformation on the attention scores. The arrays used by the PEs are tiled along both dimensions. Thus, this engine is accessed $TS_{FFN} \times TS_{FFN}$ times to finish the complete operation. The second for loop of the HLS code is pipelined causing the innermost for loop to be fully unrolled. This generates $TS_{FFN}$ PEs which equals to $\frac{d_{model}}{Tile\ no.\ FFN}$.
%\vspace{0.2cm}
\subsubsection{\textbf{{FFN2}\textsubscript{CE} engine}}
$FFN2_{CE}$ engine performs second linear transformation on the normalized outputs of $FFN1_{CE}$ engine. The arrays used by the PEs are tiled along both dimensions. Thus, this engine is accessed $4\times TS_{FFN} \times TS_{FFN}$ times to finish the complete operation. This engine also contains $TS_{FFN}$ PEs which equals to $\frac{d_{model}}{Tile\ no.\ FFN}$, because the trip count of the innermost loop is $\frac{d_{model}}{Tile\ no.\ FFN}$ and it is fully unrolled. 
%\vspace{0.2cm}
\subsubsection{\textbf{{FFN3}\textsubscript{CE} engine}}
$FFN3_{CE}$ engine performs final linear transformation on the normalized outputs of $FFN2_{CE}$ engine. The arrays used by the PEs are tiled along both dimensions. Thus, this engine is accessed $4\times TS_{FFN} \times TS_{FFN}$ times to finish the complete operation. The complete unroll of the innermost loop generates $4\times TS_{FFN}$ PEs in it, which equals to $\frac{4\times d_{model}}{Tile\ no.\ FFN}$.

\begin{algorithm}[H]
\centering
\caption{FFN Computation Example}\label{FFN1}
\begin{algorithmic}[1]
%\State Inputs
\For{$i \gets 1$ to $Sequence\ Length$}
\State \textbf{\#pragma HLS pipeline off}
\State $m \gets index \times \frac{Embedding\ Dimension}{Tile\ no.\ FFN}$
\For{$j \gets 1$ to $\frac{Embedding\ Dimension}{Tile\ no.\ FFN}$}
\State \textbf{\#pragma HLS pipeline II = 1}
\State $sum \gets 0$
\For{$k \gets 1$ to $\frac{Embedding\ Dimension}{Tile\ no.\ FFN}$}
    \State $sum \gets sum + inputs[i][k]\times weights[k][j];$
\EndFor
    \State $output[i][m] \gets output[i][j] + sum;$
    \State $m \gets m+1; $
\EndFor
\EndFor
\end{algorithmic}
\end{algorithm}

%\vspace{-0.4cm}
\subsection{Tiling Technique}
\begin{figure}[h]
\centering
\includegraphics[height=10.5cm, width=1.0\linewidth]{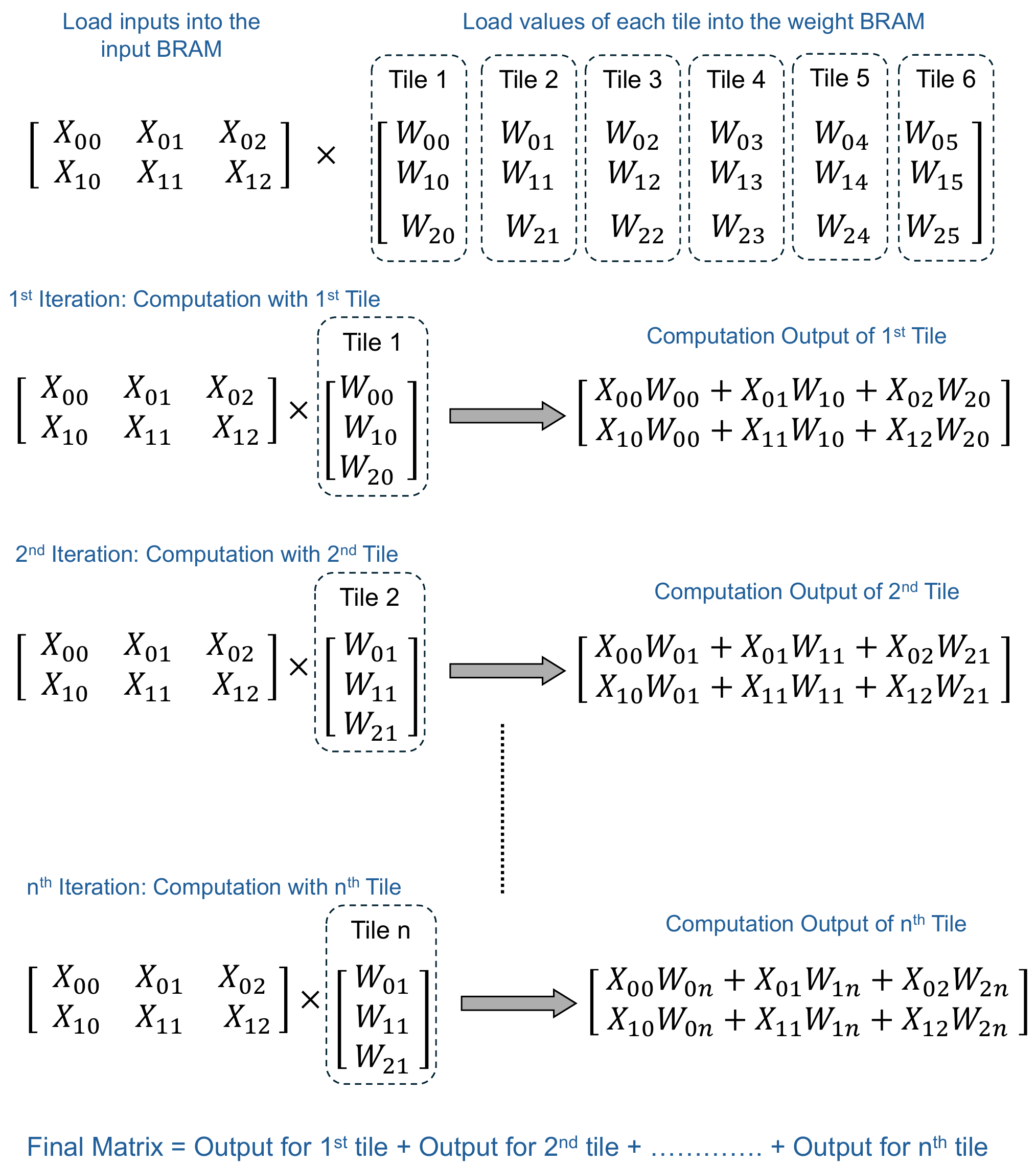} % 6cm 0.7cm
\caption{\label{tile_tech_mha}Tiling Technique in Multihead Attention Layer.}
\end{figure}

Since transformer models are typically large, tiling is used to manage the utilization of on-chip memory and computing units effectively. It ensures that the HLS tool can efficiently partition arrays and pipeline or unroll loops to minimize latency while keeping compilation time short. Figure~\ref{tile_tech_mha} illustrates our distinctive tiling strategy for the MHA module. The weight matrices are divided into tiles, enabling BRAMs to be loaded with partial data fetched from off-chip memory. Tiling is applied only along the second dimension (columns) of the matrix because the first dimension (rows) is already reduced by the number of heads. Consequently, each matrix is loaded ($\frac{d_{model}}{TS_{MHA}}$) times. The input buffers for each attention head are defined as a two-dimensional array of size (SL $\times$ $TS_{MHA}$), and tiling is similarly applied along the column of the matrix, resulting in ($\frac{d_{model}}{TS_{MHA}}$) loads. During each iteration, data for one tile is loaded initially. The PEs then compute on this data, storing the results in intermediate buffers, which are accumulated with results from previous iterations. Ultimately, the final output is the cumulative sum of the results computed across all tiles.

%\

\begin{figure}[h]
\centering
\includegraphics[height=7cm, width=1.0\linewidth]{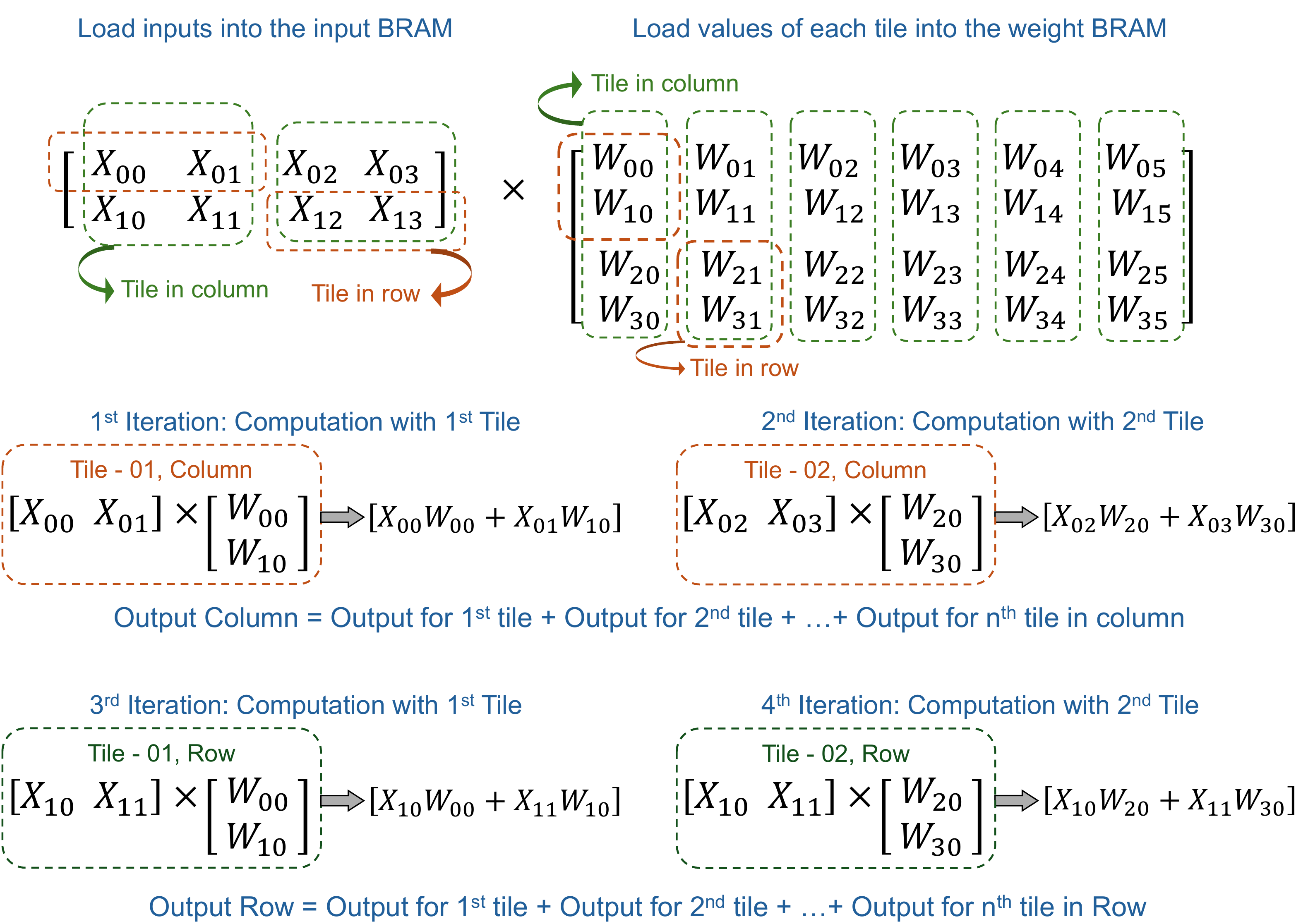} % 6cm 0.7cm
\caption{\label{tile_tech_FFN}Tiling Technique in FFN.}
\end{figure}

The FFNs that follow the attention layer are the most time- and resource-intensive components. The weight matrices for the FFN are defined as two-dimensional arrays with dimensions $(TS_{FFN}) \times (4 \times TS_{FFN})$. These matrices are tiled along both dimensions (rows and columns), requiring two loops to iteratively load each tile. The first FFN module is reused $(\frac{d_{model}}{TS_{FFN}})^2$ times because both loops iterate $\frac{d_{model}}{TS_{FFN}}$ times. The second and third FFN modules are reused $(\frac{4 \times (d_{model})^2}{(TS_{FFN})^2})$ times, reflecting the iteration counts of either $\frac{d_{model}}{TS_{FFN}}$ or $\frac{4 \times d_{model}}{TS_{FFN}}$. Figure~\ref{tile_tech_FFN} illustrates our specific tiling strategy for the FFN. Results are first accumulated along the columns, followed by accumulation along the rows for all tiles.

\vspace{-0.15cm}
\subsection{Runtime Configurable Capability}
The runtime-programmable parameters such as the number of attention heads, number of layers, embedding dimension, and sequence length can be sent to \textbf{\textit{ProTEA}} via software running on the $\mu$B processor. TNN models are trained using the PyTorch framework, and the resulting models should be saved as \textit{'.pth'} files. These files are then processed by a Python interpreter to extract key parameters such as the number of attention heads, layers, embedding dimension, and sequence length. While these parameters will vary across applications, \textbf{\textit{ProTEA}} does not require resynthesis for each model; only minor software modifications are necessary. The software, developed in C++ using the Xilinx SDK tool, utilizes the extracted data to generate instructions and control signals. These signals guide the processor in activating the relevant parts of the accelerator hardware. % using the steps shown in Fig.~\ref{isa}. , which runs on the processor

\subsection{Tile Size Determination}
\begin{figure}[h]
\centering
\includegraphics[height=5.5cm, width=1.0\linewidth]{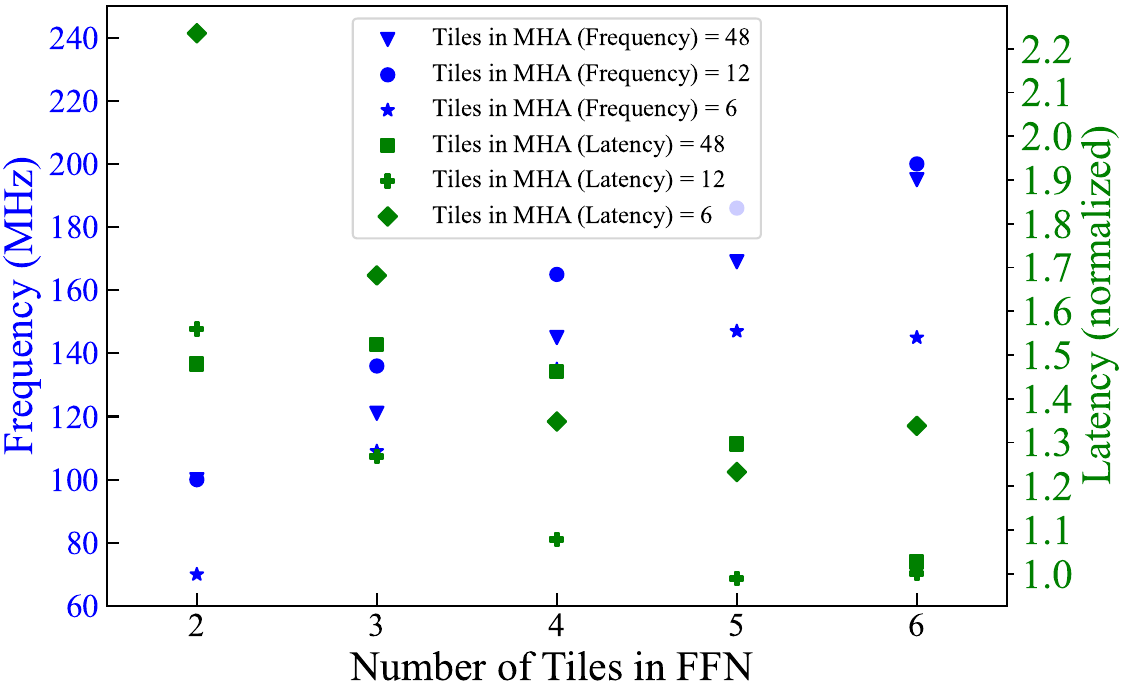} % 6cm 0.7cm
\caption{\label{tile_select}Choosing the optimum tile size.}
\end{figure}

In \textbf{\textit{ProTEA}}, the programmable parameters can be adjusted at runtime, whereas the tile size must be set before synthesis, as it cannot be modified without resynthesizing the entire hardware. The graph in Fig.~\ref{tile_select} illustrates how variations in $TS_{MHA}$ and $TS_{FFN}$ impact system frequency (MHz) and latency (normalized to the minimum value). The number of tiles in MHA ($\frac{d_{model}}{TS_{MHA}}$) was varied from 6 to 48, and for each MHA tile count, the number of tiles in FFN ($\frac{d_{model}}{TS_{FFN}}$) ranged from 2 to 6. The results indicate that the optimal configuration for achieving the highest frequency (blue color) and lowest latency (green color) was 12 tiles in MHA and 6 tiles in FFN. This setup achieved a maximum frequency of 200 MHz, allowing \textbf{\textit{ProTEA}} to execute all transformer neural network models discussed in Section~\ref{results}. Moreover, experiments showed that $TS_{MHA}$ of 64 and $TS_{FFN}$ of 128 are optimal for HLS, allowing for efficient array partitioning within a reasonable compilation time (approximately 36 hours) for a state-of-the-art (SOTA) transformer encoder.

%\vspace{-0.4cm}

\begin{table*}[h]
\setlength{\arrayrulewidth}{0.05pt}%{0.1mm}
\renewcommand{\arraystretch}{1.0}
  \centering
  \caption{Overall Results for Our Accelerator.}
  \resizebox{0.8\linewidth}{2.7cm}{%3.0cm 0.9cm %c
    \begin{tabular}{|c|c|c|c|c|c|c|c|c|c|c|c|c|}
    \hline
    %\noalign{\hrule height 2pt} !{\vrule width 2pt}
   \multicolumn{1}{|c|}{\multirow{2}[2]{*}{\textbf{Test no.}}} & \textbf{Sequence} & \textbf{Embedding } & \multicolumn{1}{c|}{\textbf{Number}} & \textbf{Number} & \textbf{Data} & \multirow{2}[2]{*}{\textbf{DSPs}}  & \multirow{2}[2]{*}{\textbf{LUTs}} & \multirow{2}[2]{*}{\textbf{FFs}}& \textbf{Latency} & \multicolumn{1}{|c|}{\multirow{2}[2]{*}{\textbf{GOPS}}} \bigstrut[t]\\
   %!{\vrule width 2pt} !{\vrule width 2pt}  & \textbf{Frequency} 
   \multicolumn{1}{|c|}{ }   & \textbf{ Length} & \textbf{Dimension} & \multicolumn{1}{c|}{\textbf{of Heads}} & \textbf{of Layers} & \textbf{Format} &   &   &   & \textbf{ (ms)} &  \multicolumn{1}{|c|}{}\bigstrut[b]\\
    %\hline !{\vrule width 2pt} & \textbf{ (MHz)} & \textbf{BRAMs}
    \noalign{\hrule height 2pt}
    \#1 & \multirow{3}[6]{*}{64} & \multirow{3}[6]{*}{768} & \multicolumn{1}{c|}{8} & \multirow{3}[6]{*}{12} &  \multirow{3}[6]{*}{8bit fixed} & \multirow{3}[6]{*}{3612 (40\%)} & \multirow{3}[6]{*}{993107 (76\%)} & \multirow{3}[6]{*}{704115 (27\%)}  & 279 & 53 \bigstrut\\ % & \multirow{3}[6]{*}{242 (6\%)}
\cline{1-1}\cline{4-4}\cline{10-11}    \#2 &   &   & \multicolumn{1}{c|}{4} &   &   &  &   &   &  285 & 51 \bigstrut\\ %& \multirow{3}[6]{*}{400} &
\cline{1-1}\cline{4-4}\cline{10-11}    \#3 &   &   & \multicolumn{1}{c|}{2} &   &   &   &   &   &  295 & 49 \bigstrut\\ %  & 
    \hline
    \multicolumn{11}{c}{} \bigstrut\\ [-1.0em]
    \hline
       \#4 & \multirow{2}[4]{*}{64} & \multirow{2}[4]{*}{768} & \multirow{2}[4]{*}{8} & \multicolumn{1}{c|}{8}  &  \multirow{2}[4]{*}{8bit fixed} & \multirow{2}[4]{*}{3612 (40\%)}  & \multirow{2}[4]{*}{993107 (76\%)} & \multirow{2}[4]{*}{704115 (27\%)}  & 186 & 80 \bigstrut\\ % & \multirow{2}[4]{*}{242 (6\%)}
\cline{1-1}\cline{5-5}\cline{10-11}    \#5 &   &   &  &  \multicolumn{1}{c|}{4} &   &  &  &   &  93 & 159 \bigstrut\\ %& \multirow{3}[6]{*}{400} &
    \hline
    \multicolumn{11}{c}{} \bigstrut\\ [-1.0em]
    \hline
    \#6 & \multirow{2}[4]{*}{64} & 512 & \multicolumn{1}{c|}{\multirow{2}[4]{*}{8}} & \multirow{2}[4]{*}{12} &  \multirow{2}[4]{*}{8bit fixed} & \multirow{2}[4]{*}{3612 (40\%)} & \multirow{2}[4]{*}{993107 (76\%)} & \multirow{2}[4]{*}{704115 (27\%)} & 186 & 36 \bigstrut\\ %& \multirow{2}[4]{*}{400} % & \multirow{2}[4]{*}{242 (6\%)} 
\cline{1-1}\cline{3-3}\cline{10-11}    \#7 &   & 256 & \multicolumn{1}{c|}{} &  &   &   &   &   &  95 & 18 \bigstrut\\
    \hline
    \multicolumn{11}{c}{} \bigstrut\\ [-1.0em]
    \hline
    \#8 & 128 & \multirow{2}[4]{*}{768} & \multicolumn{1}{c|}{\multirow{2}[4]{*}{8}} & \multirow{2}[4]{*}{12} &  \multirow{2}[4]{*}{8bit fixed} & \multirow{2}[4]{*}{3612 (40\%)} & \multirow{2}[4]{*}{993107 (76\%)} & \multirow{2}[4]{*}{704115 (27\%)} &  560 & 54 \bigstrut\\ %& \multirow{3}[6]{*}{400}  & \multirow{2}[4]{*}{242 (6\%)}
\cline{1-1}\cline{2-2}\cline{10-11}    \#9 & 32 &   & \multicolumn{1}{c|}{} &   & \multirow{2}[2]{*} &   &   &   &  165 & 44 \bigstrut\\ %& 
%\cline{1-1}\cline{2-2}\cline{11-12}    \#10 & 16 &   & \multicolumn{1}{c|}{} &   &   &   &   &   &   &  13 & %16 \bigstrut\\ % & 
    \hline
%    \multicolumn{13}{c}{} \bigstrut\\ [-1.0em]
%    \hline
%    \#9 & \multirow{2}[4]{*}{64} & \multirow{2}[4]{*}{768} & \multicolumn{1}{c|}{\multirow{2}[4]{*}{8}} & 32 & Alveo & \multirow{2}[4]{*}{8bit fixed} & 3636 (40\%) & 2636 (65\%) & 746769 (57\%) & 587337 (22\%) & 1.155 & 267 \bigstrut\\ %& \multirow{2}[4]{*}{400}
%\cline{1-1}\cline{5-5}\cline{8-11}\cline{12-13}    \#10 &   &   & \multicolumn{1}{c|}{} & 16 &  U55C &   & 2996 (33\%) & 2380 (59\%) & 607554 (46\%) & 529543 (20\%) &  1.563 & 197 \bigstrut\\ %& 
%    \hline
%    \multicolumn{13}{c}{} \bigstrut\\ [-1.0em]
%    \hline
%    \#11 & \multirow{2}[4]{*}{64} & 768 & \multirow{2}[4]{*}{6} & \multirow{2}[4]{*}{64} & Alveo  & \multirow{2}[4]{*}{8bit fixed} & \multirow{2}[4]{*}{3306 (48\%)} & \multirow{2}[4]{*}{2740 (63\%)} & \multirow{2}[4]{*}{1048022 (88\%)} & \multirow{2}[4]{*}{625983 (26\%)}  & 0.977 & 315 \bigstrut\\ % & \multirow{2}[4]{*}{500}
%\cline{1-1}\cline{3-3}\cline{12-13}    \#12 &   & 512 &   &   & U200 &   &   &   &   &   &  0.604  &  182 %\bigstrut\\ 
%    \hline
    \end{tabular}%
    }
  \label{results-overall}%
\end{table*}

%\vspace{-0.4cm}

% Table generated by Excel2LaTeX from sheet 'Sheet4'
\begin{table*}[t]%[htbp]
\setlength{\arrayrulewidth}{0.05pt}%{0.1mm}
\renewcommand{\arraystretch}{1}
  \centering
  \caption{Comparison with FPGA Accelerators.}
    \resizebox{0.6\linewidth}{2.8cm}{
    \begin{tabular}{|c|c|c|c|c|c|c|c|c|}
    \hline
    \multicolumn{1}{|c|}{\multirow{2}[2]{*}{\textbf{Accelerator}}} & \multicolumn{1}{c|}{\multirow{2}[2]{*}{\textbf{Precision}}} & \multirow{2}[2]{*}{\textbf{FPGA}} & \multirow{2}[2]{*}{\textbf{ \ DSP \ }} & \textbf{Latency } & \multirow{2}[2]{*}{\textbf{GOPS}} & \textbf{(GOPS/DSP)$\times$} & \multirow{2}[2]{*}{\textbf{Method}} & \multirow{2}[2]{*}{\textbf{Sparsity}} \bigstrut[t]\\
    \multicolumn{1}{|c|}{} & \multicolumn{1}{c|}{} &   &   & \textbf{(ms)} &   & \textbf{1000} &   &  \bigstrut[b]\\
    \hline
    \multicolumn{1}{|c|}{\cite{peng_accelerating_2021}} & -- & Alveo U200 & 3368 & 0.32 & 555 & 164 & \multirow{2}[3]{*}{HLS} & 90\% \bigstrut\\
\cline{1-7}\cline{9-9}    \textit{ProTEA} & Fix8 & Alveo U55C & 3612 & 4.48 & 79 & 22 &   & 0\% \bigstrut \\ %2.24
    \hline
    \multicolumn{9}{c}{} \bigstrut\\ [-1.3em]
    \hline
    \multicolumn{1}{|c|}{\cite {wojcicki_accelerating_2022}} & \multicolumn{1}{c|}{Float32} & Alveo 250 & 4351 & 1.2 & 0.0006 & 0.00013 & \multirow{2}[4]{*}{HLS} & \multirow{2}[4]{*}{0\%} \bigstrut\\
\cline{1-7}   \textit{ProTEA} & Fix8 & Alveo U55C & 3612 & 0.425 & 0.0017 & 0.00045 &   &  \bigstrut\\
    \hline
    \multicolumn{9}{c}{} \bigstrut\\ [-1.3em]
    \hline
     \multicolumn{1}{|c|}{\cite{yang_efa-trans_2022}} & Int8 & ZCU102 & 1024 & 1.47 & 279 & 272 & HDL & \multirow{2}[4]{*}{0\%} \bigstrut\\
\cline{1-7}\cline{8-8}  \textit{ProTEA} & Fix8 & Alveo U55C & 3612 & 5.18 & 83 & 23 & HLS &   \bigstrut\\
    \hline
    \multicolumn{9}{c}{} \bigstrut\\ [-1.3em]
    \hline
    \multicolumn{1}{|c|}{\cite{qi_accelerating_2021}} & \multicolumn{1}{c|}{--} & Alveo 200 & 4145 & 15.8 & 75.94 & 18 & \multirow{2}[4]{*}{HLS} & \multirow{2}[4]{*}{0\%} \bigstrut\\
\cline{1-7}   \textit{ProTEA} & Fix8 & Alveo U55C & 3612 & 9.12 & 132 & 37 &   &  \bigstrut\\ 
    \hline
    \multicolumn{9}{c}{} \bigstrut\\ [-1.3em]
    \hline
    \multicolumn{1}{|c|}{\cite{li_ftrans_2020}} & Fix16 & VCU118 & 5647 & 2.94 & 60 & 11 & \multirow{2}[4]{*}{HLS} & 93\% \bigstrut\\
\cline{1-7}\cline{9-9}    \textit{ProTEA} & Fix8 & Alveo U55C & 3612 & 4.48 & 79 & 22 &   & 0\% \bigstrut\\ % 2.24
    \hline
    %\multicolumn{9}{c}{} \bigstrut\\ [-1.3em]
    %\hline

\hline
    \end{tabular}%
    }
  \label{fpga}%
\end{table*}%

%\vspace{-0.6cm}
%\input{hls}
%\input{hdl}
%\input{system}
%\input{implementations}
%\vspace{-0.4cm}
\section{Evaluation and Results}\label{results}

Table~\ref{results-overall} presents the runtime programmability, resource utilization, and performance metrics of \textbf{\textit{ProTEA}}. The reported latency reflects the computation time, accounting for the overlap of data loading and computation. The synthesis was conducted with fixed tile sizes of $TS_{MHA}$ = 64 and $TS_{FFN}$ = 128, as these values are set before synthesis and cannot be altered afterward. Data was quantized to 8-bit fixed-point format; while this might result in accuracy loss depending on the application, it was not a primary focus. For applications requiring a larger bit width, the design can be easily modified in the HLS code, which will impact both resource utilization and latency. The accelerator’s design parameters, including the embedding dimension ($d_{model}$), number of heads (h), number of layers (N), and sequence length (SL), were initially configured with fixed values — 768, 8, 12, and 64 respectively — based on a variant of BERT\cite{bert} and the available FPGA resources. These parameters were then adjusted dynamically at runtime using $\mu$B. This approach allows \textbf{\textit{ProTEA}} to be synthesized once for a fixed set of resources while retaining the flexibility to adapt to various architectures as needed.

Tests 1, 2, and 3 demonstrate how varying the number of attention heads within the same accelerator dynamically impacts latency and throughput, with throughput defined as the number of giga operations per second (GOPS). On the Alveo U55C, the lowest latency of 279 ms and the highest GOPS of 53 were achieved with 8 parallel heads. Tests 4 and 5 explore the effect of varying the number of layers, showing that latency decreases and GOPS increases as the number of layers is reduced. Tests 6 and 7 examine the impact of embedding dimensions, with latency increasing and GOPS decreasing as the embedding dimension grows. Finally, Tests 8 and 9 investigate the effect of varying sequence length, where performance deteriorates as sequence length increases.

Resource utilization remained consistent across Tests 1 to 9, as the accelerator was synthesized only once with a fixed tile size, while other parameters were reconfigured at runtime through software. The design achieved high resource utilization, with 40\% of DSPs and 76\% of LUTs in use. Further DSP utilization was limited by the available LUTs, and the optimal number of parallel attention heads was determined to be 8 on the Alveo U55C to avoid overutilization by the ${QKV}_{CE}$ engine. %Reducing the tile size helped to decrease resource consumption, although at the expense of speed. Six parallel attention heads were feasible on U200, and this decrease in parallelism led to an increase in latency. %Tests no. 9 \& 10 had different tile sizes, necessitating resynthesis of the accelerator, which resulted in different resource utilization. Resource utilization decreased with a reduction in tile size, leading to increased latency and decreased GOPS. This is because a smaller tile size requires more frequent loading of each tile from external memory to on-chip memory. Tests no. 11 and 12 demonstrated the performance and resource utilization of FAMOUS on Alveo U200, highlighting its portability. 

Table~\ref{fpga} compares the performance of our accelerator, \textbf{\textit{ProTEA}}, with other FPGA-based accelerators. Each of these accelerators is custom-built for a specific TNN model, with some designed specifically for sparse computations. Among them, only EFA-Trans \cite{yang_efa-trans_2022} is flexible enough to toggle the sparse preprocessing unit, allowing it to switch between sparse and dense computations. Since \textbf{\textit{ProTEA}} was synthesized only once with a fixed set of hardware resources and bit width, and was implemented on a different platform, we evaluated performance metrics like latency, throughput (GOPS), and normalized throughput (GOPS per DSP) \cite{lstm_high_rate} for a fair comparison. \textbf{\textit{ProTEA}} achieved 2.8$\times$ and 1.7$\times$ improvements in speed and GOPS, respectively, compared to the accelerators proposed by Wojcicki et al. \cite{wojcicki_accelerating_2022} and Qi et al. \cite{qi_accelerating_2021}. The GOPS/DSP ratio was also increased by 3.46$\times$ and 2$\times$ compared to these accelerators. On the other hand, EFA-Trans, which appears to be custom-designed using HDL methods, resulted in more efficient hardware with a lower level of abstraction, making it 3.5$\times$ faster than \textbf{\textit{ProTEA}}. Peng et al. \cite{peng_accelerating_2021} applied a high sparsity of 90\% to their model, achieving a 14$\times$ speedup over \textbf{\textit{ProTEA}}. If the same sparsity level were applied to \textit{ProTEA}, its latency would mathematically be reduced to 0.448 ms (calculated as $4.48 - 4.48 \times 0.9$), making it 1.4$\times$ slower. FTRANS \cite{li_ftrans_2020} compressed the model by 93\%. The same compression would make \textbf{\textit{ProTEA}} 9.4$\times$ faster because its latency would be 0.31 ms (calculated as $4.48 - 4.48 \times 0.93$). Moreover, \textbf{\textit{ProTEA}} demonstrated 2$\times$ higher GOPS/DSP than FTRANS, indicating more efficient DSP usage.
%Peng et al. \cite{peng_accelerating_2021} applied a high sparsity of 90\% to their model, achieving a 7$\times$ speedup over \textit{ProTEA}. If the same level of sparsity were applied to \textit{ProTEA}, its latency would mathematically be reduced to 0.224 ms (calculated as $2.24 - 2.24 \times 0.9$), making it 1.42$\times$ faster. Although FTRANS \cite{li_ftrans_2020} compressed the model by 93\%, it was still 1.3$\times$ slower than \textit{ProTEA}. Moreover, \textit{ProTEA} demonstrated 2$\times$ higher GOPS/DSP than FTRANS, indicating more efficient DSP usage.

Table~\ref{compare} compares \textbf{\textit{ProTEA}} with various GPUs and CPUs operating at frequencies between 1.3 and 3.2 GHz. \textbf{\textit{ProTEA}} was tested with different TNN models, as referenced in the second column. We could easily adjust the embedding dimensions, number of heads \& layers, and sequence length in runtime to align with the architectures in the referenced studies without altering the hardware, thus, ensuring a fair comparison. \textbf{\textit{ProTEA}} is 0.79$\times$ and 6.65$\times$ slower than the Intel I5-5257U CPU and JETSON TX2 GPU respectively for model \#1 because this study \cite{peng_accelerating_2021} applied a pruning technique. It is 2.5$\times$ faster than the NVIDIA TITAN XP GPU for model \#2, and 16$\times$ faster than the NVIDIA TITAN XP GPU for model \#4. These improvements are attributed to higher parallelism, despite \textbf{\textit{ProTEA}} operating at a lower frequency and lacking sparsity. For model \#3, \textbf{\textit{ProTEA}} performed slower than the Intel I5-4460 CPU and NVIDIA RTX 3060 GPU, potentially due to the use of aggressive sparsity and omission of certain computations in the referenced work.
\vspace{-0.4cm}
\begin{table}[h]
\setlength{\arrayrulewidth}{0.05pt}%{0.1mm}
\renewcommand{\arraystretch}{0.9}
  \centering
  \caption{Cross-Platform Comparison}
    \resizebox{1.0\linewidth}{2.0cm}{
    \begin{tabular}{|c|c|c|c|c|c|}
    \hline
    \textbf{TNNs} & \textbf{Works} & \textbf{Platform} &  \textbf{Frequency} & \textbf{Latency (ms)} & \textbf{Speed Up} \bigstrut\\
    \hline
    \multirow{3}[2]{*}{\#1} & \multirow{3}[2]{*}{\cite{peng_accelerating_2021}} & INTEL I5-5257U CPU  &  2.7  GHz & 3.54 (Base) & 1 \bigstrut[t]\\
    &  & JETSON TX2 GPU  &  1.3 GHz & 0.673 & 5.3$\times$ \\
    &  & \textit{ProTEA} FPGA  & 0.2 GHz & 4.48 & 0.79$\times$ \bigstrut[b]\\
    \hline
        \multicolumn{6}{c}{} \bigstrut\\ [-1.3em]
    \hline
    \multirow{2}[2]{*}{\#2} & \multirow{2}[2]{*}{\cite{wojcicki_accelerating_2022}} & NVIDIA TITAN XP GPU  &  1.4 GHz & 1.062 (Base) & 1 \bigstrut[t]\\
    &  & \textit{ProTEA} FPGA &  0.2 GHz & 0.425 & 2.5$\times$ \bigstrut[b]\\
    \hline
    \multicolumn{6}{c}{} \bigstrut\\ [-1.3em]
    \hline
    \multirow{3}[2]{*}{\#3} & \multirow{3}[2]{*}{\cite{yang_efa-trans_2022}} & INTEL I5-4460 CPU & 3.2 GHz & 4.66 (Base) & 1 \bigstrut[t]\\
    &  & NVIDIA RTX 3060 GPU &  1.3 GHz & 0.71 & 6.5$\times$ \\
    &  & \textit{ProTEA} FPGA & 0.2 GHz & 5.18 & 0.89$\times$ \bigstrut[b]\\
    \hline
    \multicolumn{6}{c}{} \bigstrut\\ [-1.3em]
    \hline
    \multirow{2}[2]{*}{\#4} & \multirow{2}[2]{*}{\cite{qi_accelerating_2021}} & NVIDIA TITAN XP GPU & 1.4 GHz & 147 (Base) & 1 \bigstrut[t]\\
    &  & \textit{ProTEA} FPGA & 0.2 GHz & 9.12 & 16$\times$ \bigstrut[b]\\
    \hline
    \end{tabular}%
    }
  \label{compare}%
\end{table}%
\section{Conclusion \& Future Works}\label{conclude}
In this research, we developed a flexible FPGA-based accelerator for the encoder layer of a transformer neural network (TNN) using a high-level synthesis (HLS) tool. The accelerator architecture exploits FPGA parallelism and the parallel nature of the encoder itself. On the Alveo U55C platform, resources such as BRAMs, DSPs, and LUTs were maximized to enhance parallelism and minimize latency. The accelerator supports runtime programmability, allowing it to adapt to various topologies without requiring re-synthesis. An efficient tiling technique and data loading method for weight matrices were implemented to accommodate large models in on-chip memory, while preventing the overutilization of computational resources. Experimental results show that our design outperforms some CPUs and GPUs in terms of speed and throughput despite operating at a lower frequency and lacking sparsity optimizations. Additionally, it achieved 1.3 to 2.8$\times$ speed up compared to the fastest state-of-the-art FPGA-based accelerators. Although this paper focuses solely on encoder layers, future work will extend the architecture to support both encoder and decoder layers of the transformer, using the same design principles. %maximum frequency, number of parallel attention heads, embedding dimension and tile size of 328 GOPS, 400 MHz, 8, 768 and 64 respectively, on these platforms.%To address the demands for real-time reaction, the hardware accelerator for this model was subsequently constructed on an FPGA. Even with the use of certain directives, the HLS design process produces an unmanageable circuit despite being quick and simple to modify. On the other hand, while the HDL design process is lengthy, it can still result in the intended circuit, and we were able to manage the degree of parallelism. We contrasted the two designs' performance, utility, and adaptability. Then, we analyzed the differences between the performance, utilization, and flexibility of the two design strategies. The ZCU104 platform uses the outermost loop pipelining pragma to provide the lowest latency for HLS design. The outermost loop unrolling pragma can use more resources (DSP) in HLS, but it did not achieve latency that was lower than the outermost loop pipelining pragma. High resource usage may be enabled for the HDL design. As a result, we could set up the U55C so that it can fully parallelize our LSTM model, which has the highest DSP usage. It had the lowest latency at full parallelism as a consequence. Yet, ZCU104 also outperformed U55C in HDL design at the same amount of parallelism. Our HLS and HDL designs are significantly faster than the CPU, according to experimental findings. In terms of latency, throughput, frequency, or normalized throughput, the findings further demonstrate that our approach is better than the majority of current LSTM accelerators on FPGA.

\small
\bibliographystyle{IEEEtran}
    \bibliography{IEEEabrv,tnn, tnn_base}  
\vspace{12pt}
%\color{red}
%\input{reviews}
%\input{rebuttal}

\end{document}